\documentclass[10pt,conference]{IEEEtran}
\usepackage{amssymb}
\usepackage{url}
\usepackage[figuresright]{rotating}
\begin{document}

\title{Analysis of cost-efficiency of serverless approaches}
\author{\IEEEauthorblockN{
        Nakhat Syeda, Harsh Shah, Rajvinder Singh, Suraj Jaju, Sumedha Kumar, Gourav Chhabra, Maria Spichkova
    }
    \IEEEauthorblockA{
        School of Computing Technologies, RMIT University, Melbourne, Australia}
}
\maketitle
 
\begin{abstract}
In this paper, we present a survey of research studies related to the cost-effectiveness of serverless approach and corresponding cost savings.   
We conducted a systematic literature review using Google Scholar search engine, covering the period from 2010 to 2024. We identified 34 related studies, from which we extracted 17 parameters that might influence the relative cost savings of applying the serverless approach. 
%}
\end{abstract}

% \begin{keywords} 
% Software Engineering, Serverless Approach, Software Architecture, Systematic Literature Review %}
% \end{keywords}

%===============================================
 \section{Introduction}
% \uppercase

\noindent
Serverless computing is becoming very popular over the recent decade, see e.g.,  \cite{zhang2021faster,koschel2021cloud,schleier2021serverless}. 
The serverless paradigm provides many benefits for developers as it allows focusing on the actual software applications rather than on managing the underlying infrastructure. 
It is currently applied to a diverse range of software systems, including Internet of Things platforms, data-intensive applications, applications based on machine learning (ML), etc.\cite{ali2022optimizing,benedetti2021experimental,chinamanagonda2022serverless,dave2022serverless,jarachanthan2021amps,sarroca2024mlless,yu2021gillis}. 
 
One of the advantages of this technology is potential cost-savings wrt. renting or purchasing a fixed quantity of servers. Arora et al.~\cite{arora2021determining} in their recent study analysed monthly cost differences of server-based and serverless approaches, comparing them on the basis of Amazon Web Services (AWS). The results of the study confirmed that the serverless solutions are generally more cost-effective, with the scale of the advantage being different for different services, for example:
    Amazon EC2 stacks (server-based cloud), compared to serverless Lambda functions, provide only 1\% benefit, but the monthly cost for using Amazon ECS with Fargate might be 49\% lower in cost than
server-based Amazon EC2 and
Amazon RDS.

It is crucial to have a systematic view on factors having a major influence on the cost-effectiveness of the approach. This would enable a better understanding of the following points: 
\begin{itemize}
    \item for what scenarios it might be a big advantage of using the serverless approach, 
    \item 
    for which scenarios the difference might be minimal or not in favour of this approach,
    \item 
    what factors might be optimised to provide a more cost-effective solution, e.g. what might be the optimal resource size of a serverless function, see   
    \cite{eismann2021sizeless}.
\end{itemize}

\emph{Contributions:}
We present the results of  a systematic literature review (SLR) which aim is to answer the following research question: 
\emph{RQ: What parameters might influence the re\-la\-tive cost savings of the serverless approach?}  \\
We conducted the SLR using the Google Scholar search engine, covering the period from 2010 to 2024. 
We identified 34 related studies, from which we extracted 17 parameters that might influence the relative cost savings of the serverless approach.

\emph{Outline:}
 The rest of the paper is organised as follows. 
Section~\ref{sec:related} introduces the related work. 
In Section~\ref{sec:methodology}, we discuss the methodology we applied for the SRL study. 
The analysis of the identified related studies and the parameters we extracted from them is presented in Section~\ref{sec:analysis}. 
In Section~\ref{sec:limitations}, we discuss limitations of our study and threats to its validity.  
Section~\ref{sec:summary} summarises and concludes the paper.

% %======================================================
\section{Related work}
\label{sec:related}

In this section, we discuss research studies related to our SLR. In contrast to these works, we aim to provide a comprehensive overview of parameters that might influence the re\-la\-tive cost savings of the serverless approach.

\subsection{Industrial studies}
\noindent
Hamza et al.~\cite{hamza2023understanding} conducted an empirical study to understand how organisations anticipate the costs of adopting a serverless approach. 
The authors conducted interviews with 15 experts from 8 companies involved in the migration and development of serverless systems. The study demonstrated that 
the serverless approach is considered as highly suitable for unpredictable workloads, while for some high-scale applications, it might not be a cost-effective option. The authors also introduced a taxonomy for cost-comparison of  serverless vs. traditional (server-based) cloud computing. 
Adzic and Chatley~\cite{adzic2017serverless} 
presented two industrial studies with the aim of analysing whether 
migrating an application to a serverless architecture (the Lambda deployment
architecture) might reduce hosting costs. The study demonstrated that the costs might be  reduced by 66\% -- 95\%.

Eismann et al.~\cite{eismann2021state} presented a collection of serverless applications, and introduced a systematic characterisation of serverless applications as well as analysed the overall view of the research community regarding serverless application patterns and best practices.

\subsection{Literature surveys}
Barrak et al.~\cite{barrak2022serverless} conducted a systematic mapping study of ML systems that have been applied using serverless approach.  This work covers 53 recent studies and summarises the overall trend in this research area. 
Jiang et al.~\cite{jiang2024systematic} present a systematic comparative study of serverless and serverful systems for distributed ML training. 
While conducting our study, we also observed a growing research interest in the application of a serverless approach to support ML systems. 
Le et al.~\cite{li2022serverless} presented another survey on serverless computing. 
The authors focused on the existing challenges and corresponding solutions.

Further surveys on serverless computing have been presented by Hassan et al.~\cite{hassan2021survey} and Wen et al.~ \cite{wen2023rise}. The first survey aimed to cover a broad range of questions, such as state-of-the-art contributions of the technology, current trends in its application, etc. The second survey aimed to cover serverless-related questions on performance optimisation, application migration, testing, etc. 
There were also a number of literature surveys dedicated to particular aspects or applications of serverless technology. 
Li et al.~\cite{li2021survey} presented a survey of research studies on cost optimisation in serverless cloud computing.  
Cassel et al.~\cite{cassel2022serverless} presented a survey on serverless computing for the Internet of Things. 
Ghorbian et al. \cite{ghorbian2024survey} presented a survey on the scheduling mechanisms. 

\subsection{Research Embedded in Teaching}

The systematic literature survey has been conducted as part of a research project under the initiative \emph{Research Embedded in Teaching}, proposed at RMIT University, see \cite{simic2016enhancing}, \cite{spichkova2019industry}, \cite{spichkova2017autonomous} and~\cite{young2021project}.
To encourage curiosity among Bachelor's and Master's students in Software Engineering research, we suggested incorporating research and analysis components into the projects as a bonus task.
Short research projects have been sponsored by industrial partners and focused on the topics related to a software development project (capstone project) that students complete within their final year of Bachelor's/Master's studies. 

The range of the project has been broad, covering several areas of software engineering. For example, an analysis of Agile Retrospectives and the corresponding tool support has been presented in~\cite{gaikwad2019voice} and \cite{ENASE25Spichkova}. Investigations on the applicability of computer vision technologies for automated utility meter reading have been presented in \cite{spichkova2020ICSoft} and \cite{spichkova2020comparison}. 
Automated security configuration and testing of virtual machine images has been investigated in \cite{spichkova2020vm2}.   
Usage visualisation for the AWS services has been presented in~\cite{george2020usage}, while a stack management interface to present a cloud infrastructure as stacks of cloud technologies has been introduced in~\cite{spichkova2018smi}. 
Software development projects on autonomous and social robotics systems have been presented in \cite{sun2019software} and \cite{clunne2017modelling}. In contrast to the above-mentioned projects, this work focuses on the analysis of serverless approaches. 

% \cite{spichkova2020gosecure}
% \cite{sun2018software,spichkova2018automated,christianto2018enhancing,clunne2017modelling}. 

 \begin{table*}[h!]
\centering
        \caption{Inclusion Criteria}
\label{table:Inclusion_Criteria}
\small{
 \begin{tabular} {|p{4.5cm}| p{10cm}|} 
\hline
\textbf{Inclusion Criteria (IC)} & \textbf{Explanations} \\ 
 \hline
\textbf{(IC1)} Published in refereed journals, conferences, and workshops & Due to quantity restriction and to establish credibility, this review considers refereed studies because it had a process of judgement about their methodology and quality as distinct from editorial or tutorial papers. \\ \hline
\textbf{(IC2)} Publications related to the target research field & We focused on papers reporting the analysis of cost savings of the
serverless approach, as the purpose of this review was to identify parameters that might influence the relative cost savings of the
serverless approach. \\ \hline 
\textbf{(IC3)} Published after January 2010 & Applying this criterion allowed us to cover the most up-to date studies in the research field. \\ \hline
\textbf{(IC4)} Published in the English language &  Language barrier restricts this review to consider paper writing in English only.  \\
\hline
\end{tabular}}
\end{table*}
% Published between January 2010 and December 2020

\begin{table*} [h!]
\centering
\caption{Exclusion Criteria} 
\label{table:Exclusion_Criteria}
\small{ 
 \begin{tabular} {|p{4.5cm}| p{10cm}|}
\hline
\textbf{Exclusion Criteria (EC)} & \textbf{Explanations}  \\ \hline
\textbf{(EC1)}  
Published as an editorial, industrial experience, report, tutorial or keynote papers, books, patents & Including these studies would weaken the credibility of the results.
 \\ \hline
\textbf{(EC2)} 
Duplicated studies & Duplicated studies mean that the same study appears several times in the search results - either as exactly the same publication, i.e., with the same title and content (this might occur due to formatting differences in the title of the paper), or as different publications from the same author(s) covering the same approach. Thus, we only include the latest instance to ensure the credibility of our results.\\ \hline
\textbf{(EC3)} 
Not focus explicitly on cost-effectiveness and cost savings  & We excluded studies that do not focus cost-effectiveness of serverless approach and corresponding cost savings, i.e., do not cover the question on what parameters and corresponding values might influence the relative cost savings of the
serverless approach. A high-level discussion on the advantages of the serverless approach, even when it only briefly mentions terms like  ``cost-effectiveness" and ``cost savings," does not contribute to answering the set research question.
%\emph{We also excluded studies that focus on .... }
\\  \hline
\end{tabular}
}
\end{table*}
% %======================================================
\section{Methodology}
\label{sec:methodology}
 
\noindent
We followed the method of SLR proposed by Kitchenham et al.~\cite{Kitchenham2004}, as it's a well-established approach for secondary studies in evidence-based Software Engineering. This method consists of the following phases:  
\begin{enumerate}
	\item Selection of primary literature sources,
	\item Automated search using a specified search string,
	\item Refinement of the search results using specified inclusion/exclusion criteria, and
	\item Analysis of the collected data.
\end{enumerate}

To perform an automated search, we selected the primary source of data to analyse: the search was conducted using the Google Scholar search engine\footnote{\url{https://scholar.google.com/}}, where the automated search excluded patents and citations. 
The search has been conducted using the title, abstract and keywords fields of the papers. 
 
We used the following search string: 
\begin{center}
\texttt{cost saving serverless}.
\end{center} 

We also experimented with the search string adding the terms ``effectiveness'', ``approach", etc., but as the above search string provided a broader set of identified studies related to our research question, we decided to apply it without any extensions. 
 After applying this search string, we sorted the Google Scholar entries by relevance, and auto-excluded patents and citations, which led to the identification by Google Scholar of 11,000 papers.

\begin{table*}[h!]
	\centering
	\caption{Identified set of parameters}
	\label{table:parameters}
    \small{
	\begin{tabular}{  |l|l|l|  }
		\hline 
        \textbf{ID} & Parameter & Papers 
        \\
		\hline
        \hline
P1 & Allocated amount of memory  
		 & 
		 \cite{mcgrath2017serverless}, \cite{eivy2017wary}, \cite{zhang2019mark},  
		 \cite{gunasekaran2019spock}, \cite{bortolini2019investigating}, \cite{elgamal2018costless}, \cite{reuter2020cost}, \cite{martins2020benchmarking},\\
		 && \cite{oyar2019faastest}, \cite{mahajan2019optimal}, \cite{xu2019adaptive},  \cite{back2018using}, \cite{yu2020characterizing}, \cite{jarachanthan2023acts}
		 \\ \hline
P2 & Allocated computing power  
		 &
		 \cite{mcgrath2017serverless}, \cite{eivy2017wary},  \cite{xu2019adaptive}, \cite{bortolini2019investigating}, \cite{reuter2020cost},  \cite{mahajan2019optimal},  \cite{mahmoudi2020performance}, \cite{zhang2019mark},  \cite{bolscher2019leveraging}, \cite{yu2020characterizing}
		\\ \hline
P3 & Amount of reused code   
		 &
		 \cite{oakes2018sock}, \cite{elgamal2018costless}, \cite{yu2020characterizing}
		 \\ \hline
P4 & Packaging  
		 &
		 \cite{oakes2018sock}, \cite{cai2023cost}
		\\ \hline
P5 & Number of hits  
		 &
		 \cite{eivy2017wary},  \cite{gunasekaran2019spock},  \cite{reuter2020cost}, \cite{oyar2019faastest}, \cite{bolscher2019leveraging}
		\\ \hline
P6 & Auto-scaling 
		 &
		 \cite{eivy2017wary}, \cite{martins2020benchmarking}
		\\ \hline
P7 &  Code complexity   
		 &
		 \cite{eivy2017wary}
		\\ \hline
P8 & Start type 
		 &
		 \cite{jackson2018investigation}, \cite{martins2020benchmarking}, 
         \cite{raj2024empirical},  \cite{oyar2019faastest}, \cite{xu2019adaptive}, \cite{hu2025mitigating},  \cite{mahmoudi2020performance}, \cite{muller2020traffic}, \\
		 && \cite{zhang2019mark}, \cite{yu2020characterizing}, \cite{pan2023sustainable}, \cite{xiao2024making}, \cite{shahane2022serverless}, \cite{vahidinia2022mitigating}, 
         \cite{jarachanthan2023acts}
		\\ \hline
P9 & Execution time 
		 &
		 \cite{gunasekaran2019spock}, \cite{martins2020benchmarking},
         \cite{pei2024litmus}, \cite{oyar2019faastest},  \cite{kakkar2020server},   \cite{jackson2018investigation},  
		 \cite{muller2020traffic}, \cite{bolscher2019leveraging}, \cite{shahane2022serverless}, \cite{bahga2020result}
		\\  \hline
P10 & Concurrent executions   
		 &
		 \cite{fotouhi2019function}, \cite{martins2020benchmarking}
		\\ \hline
P11 & Programming language   
		 &
		 \cite{oakes2018sock}, \cite{bortolini2019investigating}, \cite{jackson2018investigation}, \cite{martins2020benchmarking},  \cite{muller2020traffic}, \cite{back2018using}, \cite{pei2024litmus}
		\\ \hline
P12 & Location of the data    
		 &
		 \cite{carver2020wukong}
		\\ \hline
P13 &  Decentralised scheduling   
		 &
		 \cite{carver2020wukong}, \cite{ghobaei2023scheduling},  \cite{ghorbian2024survey}
		\\ \hline
P14 & Task clustering   
		 &
		 \cite{carver2020wukong}
		\\ \hline
P15 & Payload 
		 &
		 \cite{martins2020benchmarking}
		\\ \hline
P16 &  Application scale
& \cite{hamza2023understanding} 
        	\\	 \hline
P17 & Extra resources %Frequency and pattern of triggers 
		 &
		 \cite{yu2020characterizing}\\
		\hline\hline
	\end{tabular}
    }
\end{table*}

The results of the automated search were refined by the inclusion and exclusion criteria presented in Tables  \ref{table:Inclusion_Criteria} and~\ref{table:Exclusion_Criteria}.  
As the stop criterion for the analysis, we selected having three consecutive pages, where all items are irrelevant wrt. specified inclusion and exclusion criteria. We reached it on page 25 of the Google Scholar results (pages have been sorted by relevance). 
This phase was conducted in two steps: 
\begin{enumerate}
   \item %\textbf{Step 3.1:} 
   Preliminary refinement, where we analysed only the title and the abstract of the paper, based on the above criteria. As application of IC2 and especially EC3 might be limited when only the title and abstract are considered, a further step of refinement is necessary. 
   \item %\textbf{Step 3.2:} 
Detailed refinement, where the results of the first step have been further filtered by checking the actual papers wrt. IC2 and EC3. 
\end{enumerate}
 
In the last phase of our study, we analysed the selected papers to extract the parameters. 
Altogether, we identified 23 relevant papers, from which we extracted 17 
parameters that might influence the relative cost savings of the
serverless approach.

%====================================================
%\newpage 
\section{Analysis of the collected data}
\label{sec:analysis}

% \paragraph{Phase 4:} Analysis of the collected data.
\noindent
 Table~\ref{table:parameters} presents the set of 17 identified parameters along with the references to the publications from which these parameters were extracted.  We group these parameters in four sets: core resource parameters (allocated amounts of memory and computing power), code-related parameters, execution parameters, as well as data and architecture parameters. 
 
\begin{figure*}[ht!]
  \centering
  \includegraphics[scale=0.58]{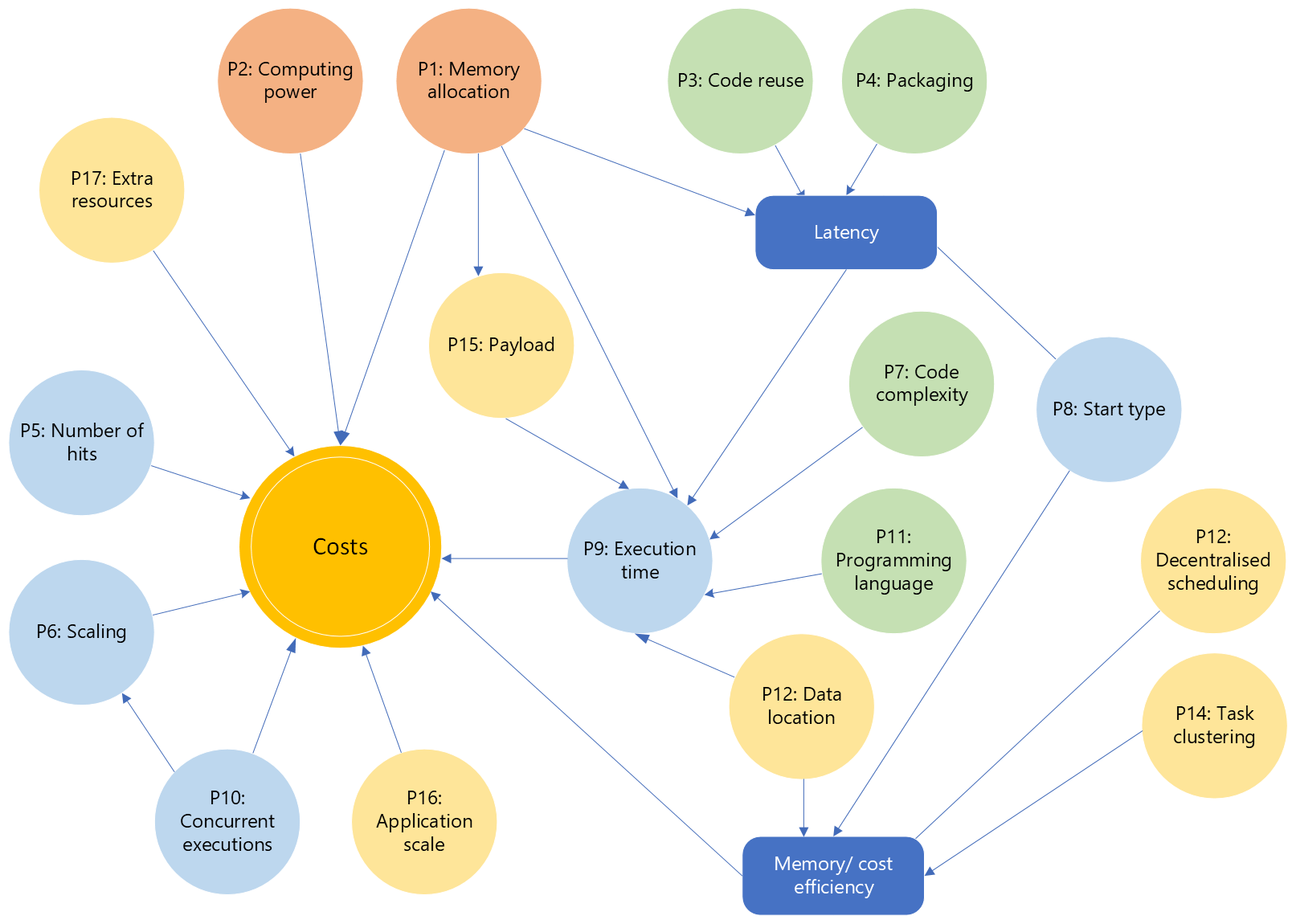}
  \caption{Correlations among parameters influencing cost-effectiveness of serverless solution}
  \label{fig:correlations}
\end{figure*}

Figure~\ref{fig:correlations} presents correlations among parameters influencing the cost-effectiveness of serverless solutions. 
In what follows, we discuss these parameters in more detail.

\subsection{Core Resource Parameters }
%~\\
\noindent 
\textbf{P1: Allocated amount of memory.} 
\textbf{P2: Allocated amount of computing power.} 
Memory allocated to a serverless function is often correlated to the amount of compute power assigned to it. 
Typically, allocation of a larger amount of memory and/or computing power means a higher cost for this allocation. However, allocation of a larger amount of memory and/or computing power might reduce the execution time of each function, see parameter P9. 
Also, having a larger amount of memory allocated helps in storing required dependencies locally, thereby reducing starting latency, which leads to a decrease in the overall execution time, see parameter P9. Therefore, to choose the right amount of memory allocated is a design trade-off, and elaborating the balance is crucial to have cost savings. 
However, over-allocation of memory and computing power might lead to a situation where these resources are underutilised/ wasted.

\subsection{Code-related  Parameters }
%~\\
\noindent
\textbf{P3: Amount of reused code.}  
Reusing code results in calling one serverless function from another increases connection latency and time taken for execution (see parameter P9), as well as increases the risk of timeout errors. 

~\\
\noindent
\textbf{P4: Packaging.}  
A larger number of packages to be installed (that are required by the code) means a larger starting latency, which leads to an increase in the overall execution time, see also parameter P9. This is especially critical for the case of the \emph{cold start}. 

~\\
\noindent
\textbf{P7:  Code complexity.}  
Code complexity influences the execution time (see parameter P9): a complex code takes longer to execute, which means higher costs to the user. 

~\\
\noindent
\textbf{P11: Programming language.}  
The language used to code the serverless function
might influence the time of execution of function, therefore affecting its pricing. Languages like Python need 10x more start-up times compared to C.

\subsection{Execution parameters}

%~\\
\noindent
\textbf{P5: Number of hits.}  
Number of hits (number of triggering events) means the
number of times a lambda function is invoked.
Serverless functions are triggered by events, making it easy to call them automatically by detecting changes in the architecture environment.
The larger the number of times a serverless function is invoked, the higher the costs to the user. 

~\\
\noindent
\textbf{P6: Auto-scaling.}  
If the execution requires auto scaling, the cost will increase. 
This might happen, e.g., in the case of a sudden increase in the concurrent requests to a function - this situation will cause auto scaling, which will increase the cost. In contrast to this,  an evenly distributed load implies lower cost due to lower scaling.

~\\
\noindent
\textbf{P8: Start type: cold vs. hot (warm).} 
For the function execution, two start types are typically considered: cold and hot (warm).  
In the case a function has to be executed for the first time or after a long period of non-execution, the execution should be set-up which involves initialisation of the execution 
environment, creating the corresponding container, etc. 
This situation is called a \emph{cold start}. 
It involves using corresponding resources and is generally more time- and resource-consuming (and, correspondingly, cost-consuming) in comparison to the case when a function has to be executed again within a short period of time, when the resources (container) can be re-used without extra initialisation and set-up. 
The second situation is called a \emph{hot (warm) start}, as the function is considered to be ``hot (warm)''. 
Thus, if an application would require 
having many cases of a cold start, this would increase overall cost of using it serverless.  

~\\
\textbf{P9: Execution time.}  
The execution time  is one of the most crucial parameters, especially in the case of lambda functions: 
There are typically time-out limits under which a function execution must be completed.  
The limit scale has a direct influence on the cost: a longer execution time means a higher cost to the user. 
In general, execution time depends on many other parameters, such as code complexity, allocated memory and computing power, etc.

~\\
\noindent
\textbf{P10: Concurrent executions.}  
Concurrent execution means the case when a function should have multiple executions simultaneously. 
As concurrent executions require scaling up, which leads to a higher cost, and therefore there is typically a limit  
to the number of concurrent executions of functions.

\subsection{Data and architecture parameters}
%~\\
\noindent
\textbf{P12: Location of the data.}  
It might play an important role in whether the data is placed locally. 
If the full data set is distributed across multiple nodes, the execution might be faster.  

~\\
\noindent
\textbf{P13:  Decentralised scheduling.}  
In decentralised scheduling, the tasks aren't coordinated by a single executor: there are multiple task executors that are responsible for the execution and scheduling of the corresponding set of tasks allocated to them. 
The advantage of decentralisation is that users aren't required to explicitly tune the number of active lambda functions running. Therefore, application of this technique makes a solution more cost-effective and more resource-efficient.

~\\
\noindent
\textbf{P14: Task clustering.}  
Task clustering means grouping some tasks that have some similarities into clusters. If this technique is applied, the set of tasks might be stored using a smaller amount of memory.

~\\
\noindent
\textbf{P15: Payload.}  
The payload of the function, i.e. its the input and output data, is strongly correlated to the required amount of memory (see P1) as well as influences the execution time (see P9). 

~\\
\noindent
\textbf{P16: Application scale.} 
The scale of the application might influence the cost-effectiveness of using a serverless approach. 
Some experts consider large-scale applications less suitable for the serverless approach.

~\\
\noindent
\textbf{P17: Extra resources.}
Another important parameter is the use of other resources required to maintain the transition of data between two stateless/serverless functions. 
This parameter outlines that the serverless is stateless; therefore, it requires the use of other services, such as storage of a permanent form like S3 or {{DynamoDB}}, which add to the costs of the architecture.

%========================================
\section{Limitations and Threats to validity}
\label{sec:limitations}

There are several threats to the validity of our systematic literature review, see \cite{zhou2016map}. In this section, we aim to identify, categorise and discuss how we attempted to mitigate them. 

\emph{Reproducibility of the study:} The research methodology has been presented in Section~\ref{sec:methodology}, describing all steps conducted in our study. This would allow us to reproduce our SLR.

\emph{Coverage of related papers:} In our study, we used the literature dataset provided by the Google Scholar. An alternative solution would be to conduct an automated search in the publisher databases directly, but as Google Scholar is one of the largest datasets covering most of the peer-reviewed publication venues, our approach allows us to obtain a reliable dataset for analysis.

\emph{Inclusion/exclusion criteria (ICs/ECs) applied for paper selection:} As individual bias and interpretation might affect the application of ICs and ECs, the authors rechecked the raw data, reviewed the application of the criteria, and suggested necessary corrections.

\emph{Accuracy of parameter extraction:} As individual bias and interpretation might affect the accuracy of parameter extractions, the consolidated list of parameters have been reviewed by the authors to reach the final consensus.

% % ===============================================
%===============================================
\section{Conclusions}
\label{sec:summary}	

\noindent 
In this paper, we present the results of a systematic literature survey. The aim of our study was to answer the research question %\\
\emph{``What parameters might influence the re\-la\-tive cost savings of the serverless approach?''}  
To answer this research question, we conducted a systematic literature review using the Google Scholar search engine, covering the period from 2010 to 2024. 
We identified 34 related studies, from which we extracted 17 parameters that might influence the relative cost savings of the serverless approach.
We analysed the extracted parameters and provided an overview of correlations among them. We also discussed the impact of the identified parameters on the the re\-la\-tive cost savings of the serverless approach.

%\vfill
\section*{\uppercase{Acknowledgements}}

We would like to thank Shine Solutions for sponsoring this project under the research grant RE-04650. 
We also would like to thank the experts from the Shine Solutions Group, especially Aaron Brown, for numerous discussions as well as their valuable feedback during the project. 

%=====================================
%\bibliographystyle{apalike}
  \bibliographystyle{plain}
%\bibliographystyle{elsarticle-num}
%\bibliography{biblio}

\begin{thebibliography}{10}

\bibitem{adzic2017serverless}
Gojko Adzic and Robert Chatley.
\newblock Serverless computing: economic and architectural impact.
\newblock In {\em 11th joint meeting on foundations of software engineering},
  pages 884--889, 2017.

\bibitem{ali2022optimizing}
Ahsan Ali, Riccardo Pinciroli, Feng Yan, and Evgenia Smirni.
\newblock Optimizing inference serving on serverless platforms.
\newblock {\em Proceedings of the VLDB Endowment}, 15(10), 2022.

\bibitem{arora2021determining}
Gary Arora, Akash Tayal, and Rakinder Sembhi.
\newblock Determining the total cost of ownership: comparing serverless and
  server-based technologies.
\newblock {\em Deloitte Consulting}, 2021.

\bibitem{back2018using}
Timon Back and Vasilios Andrikopoulos.
\newblock .using a microbenchmark to compare function as a service solutions.
\newblock In {\em European Conference on Service-Oriented and Cloud Computing},
  pages 146--160. Springer, 2018.

\bibitem{bahga2020result}
Arshdeep Bahga, Vijay~K Madisetti, Joel~R Corporan, et~al.
\newblock Result-as-a-service (raas): Persistent helper functions in a
  serverless offering.
\newblock {\em Journal of Software Engineering and Applications}, 13(10):278,
  2020.

\bibitem{barrak2022serverless}
Amine Barrak, Fabio Petrillo, and Fehmi Jaafar.
\newblock Serverless on machine learning: A systematic mapping study.
\newblock {\em IEEE Access}, 10:99337--99352, 2022.

\bibitem{benedetti2021experimental}
Priscilla Benedetti, Mauro Femminella, Gianluca Reali, and Kris Steenhaut.
\newblock Experimental analysis of the application of serverless computing to
  iot platforms.
\newblock {\em Sensors}, 21(3):928, 2021.

\bibitem{bolscher2019leveraging}
RTJ Bolscher.
\newblock Leveraging serverless cloud computing architectures: developing a
  serverless architecture design framework based on best practices utilizing
  the potential benefits of serverless computing.
\newblock Master's thesis, University of Twente, 2019.

\bibitem{bortolini2019investigating}
Diogo Bortolini and Rafael~R Obelheiro.
\newblock Investigating performance and cost in function-as-a-service
  platforms.
\newblock In {\em International Conference on P2P, Parallel, Grid, Cloud and
  Internet Computing}, pages 174--185. Springer, 2019.

\bibitem{cai2023cost}
Shen Cai, Zhi Zhou, Kongyange Zhao, and Xu~Chen.
\newblock Cost-efficient serverless inference serving with joint batching and
  multi-processing.
\newblock In {\em 14th ACM SIGOPS Asia-Pacific Workshop on Systems}, pages
  43--49, 2023.

\bibitem{carver2020wukong}
Benjamin Carver, Jingyuan Zhang, Ao~Wang, Ali Anwar, Panruo Wu, and Yue Cheng.
\newblock Wukong: a scalable and locality-enhanced framework for serverless
  parallel computing.
\newblock In {\em Proceedings of the 11th ACM Symposium on Cloud Computing},
  pages 1--15, 2020.

\bibitem{cassel2022serverless}
Gustavo Andr{\'e}~Setti Cassel, Vinicius~Facco Rodrigues, Rodrigo
  da~Rosa~Righi, Marta~Rosecler Bez, Andressa~Cruz Nepomuceno, and
  Cristiano~Andr{\'e} da~Costa.
\newblock Serverless computing for internet of things: A systematic literature
  review.
\newblock {\em Future Generation Computer Systems}, 128:299--316, 2022.

\bibitem{chinamanagonda2022serverless}
Sandeep Chinamanagonda.
\newblock Serverless data processing: Use cases and best practice-increasing
  use of serverless for data processing tasks.
\newblock {\em Innovative Computer Sciences Journal}, 8(1), 2022.

\bibitem{clunne2017modelling}
Leroy Clunne-Kiely, Bijin Idicula, Luke Payne, Enrico Ronggowarsito, Maria
  Spichkova, Milan Simic, and Heinrich Schmidt.
\newblock Modelling and implementation of humanoid robot behaviour.
\newblock {\em Procedia computer science}, 112:2249--2258, 2017.

\bibitem{dave2022serverless}
Arth Dave, Krishnateja Shiva, Pradeep Etikani, VVSR Bhaskar, and Ashok
  Choppadandi.
\newblock Serverless ai: Democratizing machine learning with cloud functions.
\newblock {\em Journal of Informatics Education and Research}, 2(1):22--35,
  2022.

\bibitem{eismann2021sizeless}
Simon Eismann, Long Bui, Johannes Grohmann, Cristina Abad, Nikolas Herbst, and
  Samuel Kounev.
\newblock Sizeless: Predicting the optimal size of serverless functions.
\newblock In {\em 22nd International Middleware Conference}, pages 248--259,
  2021.

\bibitem{eismann2021state}
Simon Eismann, Joel Scheuner, Erwin Van~Eyk, Maximilian Schwinger, Johannes
  Grohmann, Nikolas Herbst, Cristina~L Abad, and Alexandru Iosup.
\newblock The state of serverless applications: Collection, characterization,
  and community consensus.
\newblock {\em IEEE Transactions on Software Engineering}, 48(10):4152--4166,
  2021.

\bibitem{eivy2017wary}
Adam Eivy and Joe Weinman.
\newblock Be wary of the economics of"" serverless"" cloud computing.
\newblock {\em IEEE Cloud Computing}, 4(2):6--12, 2017.

\bibitem{elgamal2018costless}
Tarek Elgamal.
\newblock Costless: Optimizing cost of serverless computing through function
  fusion and placement.
\newblock In {\em 2018 IEEE/ACM Symposium on Edge Computing (SEC)}, pages
  300--312. IEEE, 2018.

\bibitem{fotouhi2019function}
Mohammadbagher Fotouhi, Derek Chen, and Wes~J Lloyd.
\newblock Function-as-a-service application service composition: Implications
  for a natural language processing application.
\newblock In {\em Proceedings of the 5th International Workshop on Serverless
  Computing}, pages 49--54, 2019.

\bibitem{gaikwad2019voice}
Purwa~Kishor Gaikwad, Chris~Theodore Jayakumar, Eashan Tilve, Niraj Bohra,
  Wenfei Yu, and Maria Spichkova.
\newblock Voice-activated solutions for agile retrospective sessions.
\newblock {\em Procedia Computer Science}, 159:2414--2423, 2019.

\bibitem{george2020usage}
Lettisia~Catherine George, Yanan Guo, Denis Stepanov, Vikas Kumar~Reddy Peri,
  Roshan~Lakmal Elvitigala, and Maria Spichkova.
\newblock Usage visualisation for the aws services.
\newblock {\em Procedia Computer Science}, 176:3710--3717, 2020.

\bibitem{ghobaei2023scheduling}
Mostafa Ghobaei-Arani and Mohsen Ghorbian.
\newblock Scheduling mechanisms in serverless computing.
\newblock In {\em Serverless Computing: Principles and Paradigms}, pages
  243--273. Springer, 2023.

\bibitem{ghorbian2024survey}
Mohsen Ghorbian, Mostafa Ghobaei-Arani, and Leila Esmaeili.
\newblock A survey on the scheduling mechanisms in serverless computing: a
  taxonomy, challenges, and trends.
\newblock {\em Cluster Computing}, pages 1--40, 2024.

\bibitem{gunasekaran2019spock}
Jashwant~Raj Gunasekaran, Prashanth Thinakaran, Mahmut~Taylan Kandemir, Bhuvan
  Urgaonkar, George Kesidis, and Chita Das.
\newblock Spock: Exploiting serverless functions for slo and cost aware
  resource procurement in public cloud.
\newblock In {\em 2019 IEEE 12th International Conference on Cloud Computing
  (CLOUD)}, pages 199--208. IEEE, 2019.

\bibitem{hamza2023understanding}
Muhammad Hamza, Muhammad~Azeem Akbar, and Rafael Capilla.
\newblock Understanding cost dynamics of serverless computing: An empirical
  study.
\newblock In {\em International Conference on Software Business}, pages
  456--470. Springer, 2023.

\bibitem{hassan2021survey}
Hassan~B Hassan, Saman~A Barakat, and Qusay~I Sarhan.
\newblock Survey on serverless computing.
\newblock {\em Journal of Cloud Computing}, 10:1--29, 2021.

\bibitem{hu2025mitigating}
Qingmiao Hu, Hongwei Li, and Elaheh Nikougoftar.
\newblock Mitigating cold start problem in serverless computing using
  predictive pre-warming with machine learning.
\newblock {\em Computing}, 107(1):1--24, 2025.

\bibitem{jackson2018investigation}
David Jackson and Gary Clynch.
\newblock An investigation of the impact of language runtime on the performance
  and cost of serverless functions.
\newblock In {\em 2018 IEEE/ACM International Conference on Utility and Cloud
  Computing Companion (UCC Companion)}, pages 154--160. IEEE, 2018.

\bibitem{jarachanthan2023acts}
Jananie Jarachanthan, Li~Chen, and Fei Xu.
\newblock Acts: autonomous cost-efficient task orchestration for serverless
  analytics.
\newblock In {\em 2023 IEEE/ACM 31st International Symposium on Quality of
  Service (IWQoS)}, pages 1--10. IEEE, 2023.

\bibitem{jarachanthan2021amps}
Jananie Jarachanthan, Li~Chen, Fei Xu, and Bo~Li.
\newblock Amps-inf: Automatic model partitioning for serverless inference with
  cost efficiency.
\newblock In {\em 50th International Conference on Parallel Processing}, pages
  1--12, 2021.

\bibitem{jiang2024systematic}
Jiawei Jiang, Shaoduo Gan, Bo~Du, Gustavo Alonso, Ana Klimovic, Ankit Singla,
  Wentao Wu, Sheng Wang, and Ce~Zhang.
\newblock A systematic evaluation of machine learning on serverless
  infrastructure.
\newblock {\em The VLDB Journal}, 33(2):425--449, 2024.

\bibitem{kakkar2020server}
Alpana Kakkar and Armaan Farshori.
\newblock Server-less cloud computing -- an economical solution for business
  operations.
\newblock In {\em Innovations in Computer Science and Engineering}, pages
  145--154. Springer, 2020.

\bibitem{Kitchenham2004}
Barbara~A. Kitchenham, Tore Dyba, and Magne Jorgensen.
\newblock Evidence-based software engineering.
\newblock In {\em Proceedings of the 26th International Conference on Software
  Engineering (ICSE)}, pages 273--281, 2004.

\bibitem{koschel2021cloud}
Arne Koschel, Samuel Klassen, Kerim Jdiya, Marc Schaaf, and Irina Astrova.
\newblock Cloud computing: serverless.
\newblock In {\em 2021 12th International Conference on Information,
  Intelligence, Systems \& Applications (IISA)}, pages 1--7. IEEE, 2021.

\bibitem{li2022serverless}
Yongkang Li, Yanying Lin, Yang Wang, Kejiang Ye, and Chengzhong Xu.
\newblock Serverless computing: state-of-the-art, challenges and opportunities.
\newblock {\em IEEE Transactions on Services Computing}, 16(2):1522--1539,
  2022.

\bibitem{li2021survey}
Zhe Li, Yusong Tan, Bao Li, Jianfeng Zhang, and Xiaochuan Wang.
\newblock A survey of cost optimization in serverless cloud computing.
\newblock In {\em Journal of Physics: Conference Series}, volume 1802, page
  032070. IOP Publishing, 2021.

\bibitem{mahajan2019optimal}
Kunal Mahajan, Daniel Figueiredo, Vishal Misra, and Dan Rubenstein.
\newblock Optimal pricing for serverless computing.
\newblock In {\em 2019 IEEE Global Communications Conference (GLOBECOM)}, pages
  1--6. IEEE, 2019.

\bibitem{mahmoudi2020performance}
Nima Mahmoudi and Hamzeh Khazaei.
\newblock Performance modeling of serverless computing platforms.
\newblock {\em IEEE Transactions on Cloud Computing}, 2020.

\bibitem{martins2020benchmarking}
Hor{\'a}cio Martins, Filipe Araujo, and Paulo~Rupino da~Cunha.
\newblock Benchmarking serverless computing platforms.
\newblock {\em Journal of Grid Computing}, pages 1--19, 2020.

\bibitem{mcgrath2017serverless}
Garrett McGrath and Paul~R Brenner.
\newblock Serverless computing: Design, implementation, and performance.
\newblock In {\em 2017 IEEE 37th International Conference on Distributed
  Computing Systems Workshops (ICDCSW)}, pages 405--410. IEEE, 2017.

\bibitem{muller2020traffic}
Lisa Muller, Christos Chrysoulas, Nikolaos Pitropakis, and Peter~J Barclay.
\newblock A traffic analysis on serverless computing based on the example of a
  file upload stream on aws lambda.
\newblock {\em Big Data and Cognitive Computing}, 4(4):38, 2020.

\bibitem{oakes2018sock}
Edward Oakes, Leon Yang, Dennis Zhou, Kevin Houck, Tyler Caraza-Harter,
  Andrea~C Arpaci-Dusseau, and Remzi~H Arpaci-Dusseau.
\newblock Sock: Serverless-optimized containers.
\newblock {\em USENIX Open Access Policy}, page~17, 2018.

\bibitem{oyar2019faastest}
Tal Oyar and Afik Deri.
\newblock Faastest-machine learning based cost and performance faas
  optimization.
\newblock In {\em Economics of Grids, Clouds, Systems, and Services: 15th
  International Conference, GECON 2018, Pisa, Italy, September 18--20, 2018,
  Proceedings}, volume 11113, page 171. Springer, 2019.

\bibitem{pan2023sustainable}
Shanxing Pan, Hongyu Zhao, Zinuo Cai, Dongmei Li, Ruhui Ma, and Haibing Guan.
\newblock Sustainable serverless computing with cold-start optimization and
  automatic workflow resource scheduling.
\newblock {\em IEEE Transactions on Sustainable Computing}, 2023.

\bibitem{pei2024litmus}
Qi~Pei, Yipeng Wang, and Seunghee Shin.
\newblock Litmus: Fair pricing for serverless computing.
\newblock {\em ASPLOS}, 2024.

\bibitem{raj2024empirical}
Vinay Raj, Himanshu Chhaparwal, and Satwik Saale.
\newblock Empirical evaluation of cold start latency in serverless platforms.
\newblock In {\em 2024 3rd International Conference for Advancement in
  Technology (ICONAT)}, pages 1--6. IEEE, 2024.

\bibitem{reuter2020cost}
Anja Reuter, Timon Back, and Vasilios Andrikopoulos.
\newblock Cost efficiency under mixed serverless and serverful deployments.
\newblock In {\em 2020 46th Euromicro Conference on Software Engineering and
  Advanced Applications (SEAA)}, pages 242--245. IEEE, 2020.

\bibitem{sarroca2024mlless}
Pablo~Gimeno Sarroca and Marc S{\'a}nchez-Artigas.
\newblock Mlless: Achieving cost efficiency in serverless machine learning
  training.
\newblock {\em Journal of Parallel and Distributed Computing}, 183:104764,
  2024.

\bibitem{schleier2021serverless}
Johann Schleier-Smith, Vikram Sreekanti, Anurag Khandelwal, Joao Carreira,
  Neeraja~J Yadwadkar, Raluca~Ada Popa, Joseph~E Gonzalez, Ion Stoica, and
  David~A Patterson.
\newblock What serverless computing is and should become: The next phase of
  cloud computing.
\newblock {\em Communications of the ACM}, 64(5):76--84, 2021.

\bibitem{shahane2022serverless}
Vishal Shahane.
\newblock Serverless computing in cloud environments: Architectural patterns,
  performance optimization strategies, and deployment best practices.
\newblock {\em Journal of AI-Assisted Scientific Discovery}, 2(1):23--43, 2022.

\bibitem{simic2016enhancing}
Milan Simic, Maria Spichkova, Hainz Schmidt, and Ian Peake.
\newblock Enhancing learning experience by collaborative industrial projects.
\newblock In {\em International Conference on Engineering Education and
  Research (ICEER)}, 2016.

\bibitem{spichkova2019industry}
Maria Spichkova.
\newblock Industry-oriented project-based learning of software engineering.
\newblock In {\em 2019 24th International conference on engineering of complex
  computer systems (ICECCS)}, pages 51--60. IEEE, 2019.

\bibitem{spichkova2018smi}
Maria Spichkova, Jesse Bartlett, Ryan Howard, Adrian Seddon, Xing Zhao, and
  Yuanqing Jiang.
\newblock Smi: Stack management interface.
\newblock In {\em 2018 23rd International Conference on Engineering of Complex
  Computer Systems (ICECCS)}, pages 156--159. IEEE, 2018.

\bibitem{ENASE25Spichkova}
Maria Spichkova, Hina Lee, Kevin Iwan, Madeleine Zwart, Yuwon Yoon, and Xiaohan
  Qin.
\newblock {Agile Retrospectives: What went well? What didn't go well? What
  should we do?}
\newblock In {\em 20th International Conference on Evaluation of Novel
  Approaches to Software Engineering (ENASE)}, 2025.

\bibitem{spichkova2020vm2}
Maria Spichkova, Biao Li, Lachlan Porter, Luke Mason, Ye~Lyu, and Yi~Weng.
\newblock Vm2: Automated security configuration and testing of virtual machine
  images.
\newblock {\em Procedia Computer Science}, 176:3610--3617, 2020.

\bibitem{spichkova2017autonomous}
Maria Spichkova and Milan Simic.
\newblock Autonomous systems research embedded in teaching.
\newblock In {\em International Conference on Intelligent Interactive
  Multimedia Systems and Services}, pages 268--277. Springer, 2017.

\bibitem{spichkova2020ICSoft}
Maria Spichkova and Johan {Van Zyl}.
\newblock Application of computer vision technologies for automated utility
  meters reading.
\newblock In {\em International Conference on Software Technologies}, pages
  521--528. INSTICC, SciTePress, 2020.

\bibitem{spichkova2020comparison}
Maria Spichkova, Johan Van~Zyl, Siddharth Sachdev, Ashish Bhardwaj, and Nirav
  Desai.
\newblock Comparison of computer vision approaches in application to the
  electricity and gas meter reading.
\newblock In {\em Evaluation of Novel Approaches to Software Engineering: 14th
  International Conference}, pages 303--318. Springer, 2020.

\bibitem{sun2019software}
Chong Sun, Jiongyan Zhang, Cong Liu, Barry Chew~Bao King, Yuwei Zhang, Matthew
  Galle, Maria Spichkova, and Milan Simic.
\newblock Software development for autonomous and social robotics systems.
\newblock In {\em Intelligent Interactive Multimedia Systems and Services:
  Proceedings of 2018 Conference 11}, pages 151--160. Springer, 2019.

\bibitem{vahidinia2022mitigating}
Parichehr Vahidinia, Bahar Farahani, and Fereidoon~Shams Aliee.
\newblock Mitigating cold start problem in serverless computing: A
  reinforcement learning approach.
\newblock {\em IEEE Internet of Things Journal}, 10(5):3917--3927, 2022.

\bibitem{wen2023rise}
Jinfeng Wen, Zhenpeng Chen, Xin Jin, and Xuanzhe Liu.
\newblock Rise of the planet of serverless computing: A systematic review.
\newblock {\em ACM Transactions on Software Engineering and Methodology},
  32(5):1--61, 2023.

\bibitem{xiao2024making}
Ke~Xiao, Song Yang, Fan Li, Liehuang Zhu, Xu~Chen, and Xiaoming Fu.
\newblock Making serverless not so cold in edge clouds: A cost-effective online
  approach.
\newblock {\em IEEE Transactions on Mobile Computing}, 2024.

\bibitem{xu2019adaptive}
Zhengjun Xu, Haitao Zhang, Xin Geng, Qiong Wu, and Huadong Ma.
\newblock Adaptive function launching acceleration in serverless computing
  platforms.
\newblock In {\em 2019 IEEE 25th International Conference on Parallel and
  Distributed Systems (ICPADS)}, pages 9--16. IEEE, 2019.

\bibitem{young2021project}
Jeffery Young, Maria Spichkova, and Milan Simic.
\newblock Project-based learning within ehealth, bioengineering and biomedical
  engineering application areas.
\newblock {\em Procedia Computer Science}, 192:4952--4961, 2021.

\bibitem{yu2021gillis}
Minchen Yu, Zhifeng Jiang, Hok~Chun Ng, Wei Wang, Ruichuan Chen, and Bo~Li.
\newblock Gillis: Serving large neural networks in serverless functions with
  automatic model partitioning.
\newblock In {\em 2021 IEEE 41st International Conference on Distributed
  Computing Systems (ICDCS)}, pages 138--148. IEEE, 2021.

\bibitem{yu2020characterizing}
Tianyi Yu, Qingyuan Liu, Dong Du, Yubin Xia, Binyu Zang, Ziqian Lu, Pingchao
  Yang, Chenggang Qin, and Haibo Chen.
\newblock Characterizing serverless platforms with serverlessbench.
\newblock In {\em Proceedings of the 11th ACM Symposium on Cloud Computing},
  pages 30--44, 2020.

\bibitem{zhang2019mark}
Chengliang Zhang, Minchen Yu, Wei Wang, and Feng Yan.
\newblock Mark: Exploiting cloud services for cost-effective, slo-aware machine
  learning inference serving.
\newblock In {\em 2019 {USENIX} Annual Technical Conference}, pages 1049--1062,
  2019.

\bibitem{zhang2021faster}
Yanqi Zhang, {\'I}{\~n}igo Goiri, Gohar~Irfan Chaudhry, Rodrigo Fonseca, Sameh
  Elnikety, Christina Delimitrou, and Ricardo Bianchini.
\newblock Faster and cheaper serverless computing on harvested resources.
\newblock In {\em Proceedings of the ACM SIGOPS 28th Symposium on Operating
  Systems Principles}, pages 724--739, 2021.

\bibitem{zhou2016map}
Xin Zhou, Yuqin Jin, He~Zhang, Shanshan Li, and Xin Huang.
\newblock A map of threats to validity of systematic literature reviews in
  software engineering.
\newblock In {\em 2016 23rd Asia-Pacific Software Engineering Conference
  (APSEC)}, pages 153--160. IEEE, 2016.

\end{thebibliography}
%\bibliographystyle{elsarticle-harv}
%{\small
%}

\end{document}